\documentclass[a4paper, 11pt] {article}

\usepackage{amssymb}
\usepackage{amsmath}
\usepackage{amscd, amssymb}
\usepackage{amsthm}
\usepackage{latexsym}
\usepackage{epsfig}
\usepackage[all]{xy} 
\usepackage{psfrag} 
\usepackage{amsfonts}
\usepackage{mathrsfs} 
\usepackage{graphicx}
\usepackage[retainorgcmds]{IEEEtrantools}
\usepackage{eurosym}

\theoremstyle{plain}

\newtheorem{thm}{Theorem} 

\newtheorem{theo}{Theorem}[section]

\newtheorem{prp}[thm]{Proposition}

\theoremstyle{definition}

\newtheorem{prop-defi}[theo]{Proposition-Definition}
\newtheorem{lemma-defi}[theo]{Lemma-Definition}

\theoremstyle{remark}

\newtheorem{ex}[theo]{Example}

\newcommand{\email}[1]{{\it{E-mail address: }} {\tt{#1}}}
\newcommand{\address}[1]{{\sc{#1}}}
\newcommand{\tn}[1]{\textnormal{#1}}
\newcommand{\ts}[1]{\textsuperscript{#1}}
\title{A Markov Chain Model for the Cure Rate of Non-Performing Loans }

\author{Vilislav Boutchaktchiev}
\newcommand{\keywords}[1]{\noindent{\bf Key words:} #1}
\newcommand{\mathsubjclass}[2][2010]{\noindent{\bf #1 Mathematics Subject
    Classification:} #2}
\newcommand{\econsubjclass}[1]{\noindent{\bf JEL Classification:} #1}

\date{1 May, 2018}

\begin{document}
\maketitle
\abstract{A Markov-chain model is developed for the purpose estimation 
  of the cure rate of non-performing loans. The technique is performed
  collectively, on  
  portfolios and it can be applicable in the process of calculation of
  credit impairment. It is  
  efficient in terms of data manipulation costs which makes it 
  accessible even to smaller financial institutions. In addition,
  several other applications to portfolio 	optimization  are
  suggested. } 
\medskip

\medskip

%{\bf 2010 Mathematics Subject Classification:} 62P20,  91B64.
\mathsubjclass{62M05, 62N02, 91B70} % 91B64.}

%{\bf JEL Classification:} G21, E51.
\econsubjclass{G21, M41}

\section*{Introduction}
In calculating the credit impairment the IFRS 9 standard permits,
under certain conditions, the usage of cure rate in order to reduce
the amount of bank provisions. The logic behind this allowance is,
that if an impaired amount will eventually return to regular status,
the bank need not calculate provisions on it. Several methodological
manuals are available to the banking community which
(cf. e.g.~\cite{basel}), often without stipulation on the assumptions
on the model, provide recipes for calculation of a cure
rate. Estimates made in this way turn out to be often overly
conservative and, sometimes, dissatisfactory, because the basic
assumptions of the model could not be verified.  The presented
technique allows to calculate  with any desired accuracy and  any
desired frequency. (E.g., monthly, 
quarterly, etc.) The model uses data only from the past 12 months in
order to provide a {\it most 
recent measurement} of the  cure rate. This is, sometimes, required by
regulators for financial 
quantities measured collectively. This very fact is the reason why in
{\it low-default} portfolios, which 
 are often those found in small banks, the results  of this type of
 method do not satisfy even very basic  
 assumptions about the cure rate in general. For this reason we apply
 a smoothing method from 
 Survival analysis. This method is the topic of Section~\ref{sec:survival}.

The usage of Markov-chain models, in general, is a
technique accessible to the banking management and is part of their
routine in accessing credit risk and expected credit losses, several
studies document and contribute to this practice, including  \cite{g-k-m},
\cite{g-a} and \cite{j-l-t}. 
It appears that  Monte Carlo techniques similar to those
demonstrated in e.g.,  \cite{v1}  
%and \cite{v2} 
are applicable in the study of cure rate and it is my belief that
future interest in this subject would move  in this direction. 

In Section~\ref{example} we produce two numerical examples using data form three
small Bulgarian banks\footnote{Bulgarian banks are bound to report
  expected credit losses and to calculate provisions based on IFRS 9
  since 01/01/2018. They are, in general, computing cure rates for
  their retail loan portfolios and for other portfolios of
  standardized products.}. 

 In addition, in Section~\ref{sec:opt} we show how this method
 provides several tools to identify 
some portfolios where cure rate is inapplicable, but rather a
different managerial approach will be more successful. 

%\section{Acknowledgment }
I wish to thank my colleagues Jana Kostova and Alessandro Merlini from
ASTOR BG Consulting for bringing this problem to my attention and for
providing the data for the actual computations.

\section{Cure Rate }\label{curerate}
Cure rate is meant to measure the propensity of loans to return to
regular status after they have been found delinquent. In a 
portfolio, collectively, the cure rate estimates what proportion of
non-performing 
loans will be, in the end, repaid. Given the possibility of a loan
which is once “cured” to relapse or to move back and forth between
categories it is not enough to simply measure the proportion over a
certain horizon of time.  

For the purposes of this study we assume that a loan is considered
non-performing after it is found more than 90 days late. We make the
following assumptions 
\begin{enumerate}
\item The loan is finally cured after it becomes less than one month
  past due. 
\item Loans which are N or more months past due are considered lost
  and are written off.  
\item We should distinguish performing loans which have been granted
  forbearance. According to \cite{aqr} these would be loans to parties
  experiencing financial difficulties in meeting their
  obligations and the bank has agreed to offer them special
  contractual terms. If such loan preserves its regular status for a
  year we consider it cured, otherwise we consider it lost.
\item States are assigned to all loans in the portfolio, based on $m$,
  the whole number of months past-due at time $t=0$. The 
  state where $m=0$ is an absorbing state, as well as the one with
  $m\ge N$.  The forborne loans are assigned in a separate
  state. Hence, the number of states is $N+2$. 
\item We assume the time periodicity of observation to the loan tape
  is annual. We measure the probabilities of transition between states
  by observing the migration between states within a year prior to
  time $t = 0$. 
\item We assume that the migration in the previous years is
  irrelevant to the further development of the portfolio. Moreover, we
  assume that transition rates do not vary in time. 
\end{enumerate}
Assumptions 1-3 are a question of bank policy and, although they
satisfy the requirement of IFRS 9, an alternative configuration may be
set.  Assumption 5 is inessential, although it should be noted that
small banks often have shallow, low-default portfolios and high
frequency observation leads to volatile cure rates. 

\section{Markov Process}
For a typical loan of the considered portfolio we have thus
constructed a 
finite Markov chain  of random variables
$\{X_t:t=0,1,...\}$, which take as value the state of the loan at
year $t$.
It has $N+2$ states 
$\{S_i: i=0, \dots, N+1\}$, describing the state the loan is.
With an appropriate ordering we can assume that 
$S_0$ denotes the state where $m=0$, 
$S_1$ denotes $m\ge N$,   
$S_2$ is the {\it forborne} state
and, for any $i>=3$, the state $S_i$ is characterized by
$M=i-2$. Hence, $S_5$ is the first non-performing
state, corresponding to $M=3$. 
\begin{itemize}
\item Assumption 6 form
Section~\ref{curerate} implies time homogeneity. 
\item Assumptions 1 and 2 imply that $S_0$ and $S_1$ are absorbing
  states and, hence, they form, each by itself two recurrent
  communication classes.
\item If any part of the set of states ${\cal T}= \{ S_i: i=2,\dots, N+2\}$
  forms a recurrent class this would imply that the loan contract for
  this particular portfolio can be optimized. We are giving an example
  to illustrate this in Section~\ref{example}. For this reason we
  assume that $\cal T$ is the set of  transitive states.
\item Assumption 3 implies that $S_2$ is a transitive. In fact,
$$P[X_n=S_0|X_{n-1}=S_2]=p, \quad
P[X_n=S_1|X_{n-1}=S_2]=q, 
$$
$$
P[X_n=S_i|X_{n-1}=S_2]=0, \tn{ for } i\ge 1,
$$
where $p$ and $q$, satisfying $p+q=1$ are the probabilities to survive
and fail, respectively.
\end{itemize}

We write the transition matrix
$A= (p(i,j)=P[X_1=S_j|X_0=S_i]$, therefore as follows:
$$A=\left(
  \begin{array}{c|c}
    \begin{array}{cc}
      1&0\\
      0&1
    \end{array}
  &
    \begin{array}{ccc}
      0&\dots&0\\
      0&\dots&0\\
    \end{array}\\
\hline
T
 %   \begin{array}{c}
  %    \begin{array}{cc}
  %      p&q
  %    \end{array}
  %    \\
  %    T
  %  \end{array}
    & 
    S
  \end{array}
\right).
$$
Furthermore, for the limit matrix $A_\infty= \lim_{n\to\infty} A^n$ we
have:
$$A_\infty=\left(
  \begin{array}{c|c}
    \begin{array}{cc}
      1&0\\
      0&1
    \end{array}
  &
    \begin{array}{ccc}
      0&\dots&0\\
      0&\dots&0\\
    \end{array}\\
\hline
T_\infty
%    \begin{array}{c}
%      \begin{array}{cc}
%        p&q
%      \end{array}
%      \\
%      T_\infty
%    \end{array}
    & 
    O
  \end{array}
\right),
$$
where $O$ is the zero matrix. For the matrix $T_\infty$ we have
$$T_\infty= 
\left(
  \begin{array}{cc}
   p&q\\		
    p_1&q_1\\
    \vdots&\vdots\\
    p_{N-1}&q_{N-1}
  \end{array}
\right),
$$
where $p_i$ and $q_i$ satisfy $p_i+q_i=1$ and are the probabilities of
a loan showing $i$ months of payment delay to be cured or lost,
respectively. Hence, in search for the cure rate, our goal is to
study the vector $(p_0,...p_N)$. 
\begin{prp}
  In the notation defined above, the probability to cure for a loan which is
  $i$ months past due at time $t=0$ can be found on the $i$\ts{th} row of
  the first column of the matrix
$$ T_\infty= (I-S)^{-1} T.
$$
\end{prp}
\proof
Since $S$ is a substochastic matrix, representing the transition rates
of transitive states, we know that $S^n\to O$ as $n\to\infty$. For
this reason the matrix $I-S$, with $I$ --- the identity matrix of size
$N-1$ is, indeed, invertible.

Denote by $t_{ij}$ the probability of a loan with initial state
$X_0=S_i$ to reach eventually the state $S_j$, $j=0, 1$. (In the
notation above, these are the entries of the matrix $T$, $t_{i0}=p_i$,
and $t_{i1}=q_i$.)
We have
\begin{IEEEeqnarray*}{rCl}
t_{ij}&=&P[X_n=S_j \tn{ for some } n|X_0=S_i]\\
&=&P[X_1=S_j|X_0=S_i]\\
&&+\sum_{k=0}^N P[X_n=S_j \tn{ for some }n|X_1=S_k]P[X_1=S_k|X_0=S_i
]\\
&=& p(i,j) + \sum_{k=2}^N t_{kj}p(i,k).
\end{IEEEeqnarray*}
That is,
$$ T_\infty = T + T_\infty S
$$
Hence
$$ T_\infty = (I-S)^{-1}T.
$$
\qed

\section{Survival Model}\label{sec:survival}
Let $S(x)$ be probability of a loan to be cured if the initial state
is at least $x$ months past due. For {\it non-preforming} states, $i\ge 3$, 
$$S(x)=P(X_n=S_0 \tn{ for some } n |X_0=S_i, i\ge x+2).
$$
This function needs to satisfy the following conditions:
\begin{enumerate}
\item $S(0)=1$
\item $S(x)=0$ for $x\ge N$
\item $S(x)$ is non-increasing.
\end{enumerate}
In addition, one would expect that chances of failure would increase
as a function of $x$, the months past-due. This is due, in part, to
two reasons. First, the longer delay signifies a  more dire economic status.
 And second, portfolio manager would make more effort to increase the
opportunities of survival of these loans which are less in delay, since
they have better chance. This gives us an extra condition

\begin{enumerate}
\item[4] The logarithmic derivative $\frac{1}{S(x)}\frac{dS}{dx}(x)$
  is decreasing. 
\end{enumerate} 
Generally the outcome of calculating the Markov chain need not satisfy
these conditions.  In order to smoothen the results we apply tools
form the Survival Analysis. (Cf. e.g. \cite{weibull}.) A common choice 
of survival function is a best-fitting Weibull curve:
$$S(x)=e^{-\left(\frac{x}{\lambda}\right)^k}, 
$$
corresponding to a Weibull distribution with CDF 
$ F(x) = 1- S(X)$.
Condition 4 simply means that the hazard rate is an increasing
function which would imply that the shape parameter $k$ of the curve
satisfies $k>~1$. 
The parameters $k$ and $\lambda$ must be chosen by fitting the CDF of
this distribution  
%$$f(x)=\frac{k}{\lambda}
%\left(
%  \frac{x}{\lambda}
%\right)^{k-1}
%e^{-
%  \left(
%    \frac{x}{k}
%  \right)^k}
%$$
to the points 

$$(0,1), (1, p_1), (2, p_2), \dots,  (N-1, p_{N-1}), (N, 0).
\footnote{The point $(\delta, p)$ may be added to the
  sequence with a appropriately chosen small value of the parameter
  $\delta$.}
$$

After this, the cure rate of the portfolio is equal to $S(3)$.

\section{Numerical Examples}\label{example}
\begin{ex}\label{ex:cc} {\bf A Credit-Card Portfolio}\\
We now consider the portfolio of select credit cards from a small bank
at the end of March 2007. The total size of the portfolio is
\euro328.9 Thousand, 
consisting of 1185 loans. The bank management assumes that  a loan not
serviced for 
8 or more months is lost, $N=8$. The transition matrix  $A$, according
to the 
notation above   is:

\footnotesize
$$
A=\left(
\begin{array}{rr|rrrrrrrr}
1 &0 &0 &0 &0 &0 &0 &0 &0 &0\\
0 &1 &0 &0 &0 &0 &0 &0 &0 &0\\\hline
0.37 &0.63 &0 &0 &0 &0 &0 &0 &0 &0\\ 
0.39 &0.11 &0.1 &0.157 &0.008 &0.015 &0.11 &0.06 &0.02 &0.03\\
0.37 &0.12 &0.02 &0.003 &0.012 &0.045 &0.09 &0.04 &0 &0.3\\
0.05 &0.32 &0.09 &0.004 &0.107 &0.113 &0.141 &0.102 &0.073 &0\\
0 &0.45 &0 &0 &0 &0.19 &0.119 &0.149 &0.012 &0.08\\
0 &0.4 &0 &0 &0 &0.08 &0.01 &0.31 &0 &0.2\\ 
0 &0.21 &0 &0 &0 &0.05 &0.009 &0.111 &0.41 &0.21\\
0 &0.47 &0.004 &0 &0 &0 &0 &0.037 &0.27 &0.219\\	
\end{array}
\right)
$$
\normalsize

Next, we compute the matrices $(I-S)^{-1}$: 

\footnotesize
$$
(I-S)^{-1}=
\left(
  \begin{array}{rrrrrrrr}
1 &0 &0 &0 &0 &0 &0 &0\\
0.127 &1.187 &0.018 &0.08 &0.166 &0.179 &0.121 &0.148\\
0.033 &0.004 &1.024 &0.109 &0.127 &0.172 &0.254 &0.519\\
0.114 &0.006 &0.132 &1.221 &0.216 &0.288 &0.254 &0.215\\
0.029 &0.002 &0.032 &0.299 &1.192 &0.348 &0.187 &0.274\\
0.017 &0.001 &0.018 &0.164 &0.048 &1.549 &0.237 &0.472\\
0.018 &0.001 &0.018 &0.162 &0.053 &0.396 &2.016 &0.656\\
0.012 &0 &0.007 &0.064 &0.021 &0.21 &0.708 &1.529\\
  \end{array}
\right)
$$
\normalsize
and $T_\infty$:

\footnotesize
$$
T_\infty=
\left(
  \begin{array}{rr}
0.37 &0.63\\
0.52 &0.48\\
0.398 &0.602\\
0.155 &0.845\\
0.038 &0.962\\
0.021 &0.979\\
0.021 &0.979\\
0.01 &0.99\\
  \end{array}
\right).
$$
\normalsize

Next we fit a Weibull curve on ten points:
$$
(0,  1),
(0.5,0.37), 
(1,  0.52),
(2,  0.398),
(3,  0.155 ),
$$
$$
(4,  0.038 ),
(5,  0.021),
(6,  0.021),
(7,  0.01 ),
(8, 0),
$$
using a linear regression. The results are shown in Table~\ref{tab:1}.

\begin{table}[hbtp]
  \def\arraystretch{1.4}
  \centering
     \caption{{\small{The fitting of the Weibull curve for a Credit Card
    portfolio. In
    parenthesis are 
    shown the $t$-statistic values from the linear regression. The
    ${}^*$ denotes 
    significance to the 99\% level.}}}
  \label{tab:1}
  \begin{tabular}{ccc}
\hline\hline
    $\lambda$&$k$&$R^2$\\
    \hline
    $1.51^{*}$&$1.14^*$&$0.96$\\ 
    $(2.91)$&$(15.03)$&\\
\hline\hline
  \end{tabular}
\end{table}

The shape coefficient $k=1.14$ is different from 0 at the 99\% level
of significance.  The one-sided hypothesis for $k\le 1$ is rejected at
the 95\% level\footnote{The $p$-value of the one-sided Wald test comes
  to 0.049. % 
%In fact, a large part of
%the tested portfolios of this type exhibited $k$ 
%statistically indistinguishable from 1. 
%This signifies, that there may be contractual
%change that the management may take in order to better channel the
%behavioral patterns of their  
%clients. 
}. The expected cure rate is computed as $S(3) =11.26\%$.

\begin{figure}[hbtp]
  \centering
  \includegraphics[width=.8\textwidth]{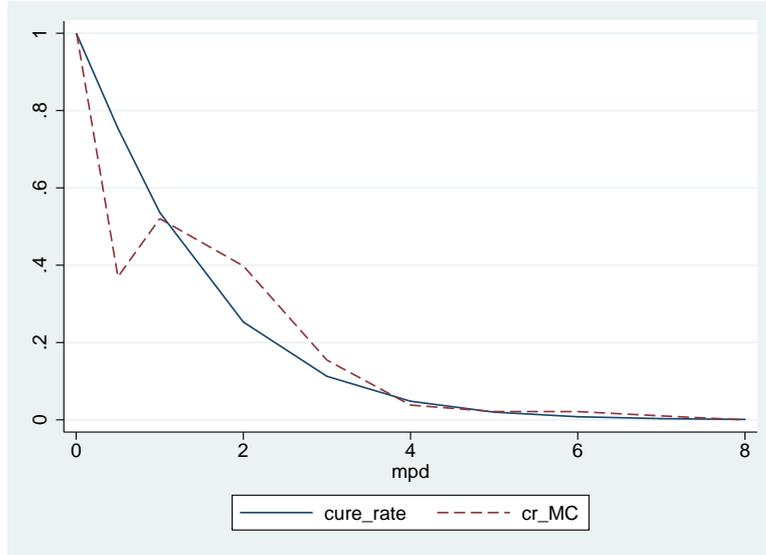} 
  \caption{{\small{The effect of Weibull curve fitting. The curve
        labeled {\it cr\_MC} represents the result of the Markov chain
        model; the final result is labeled {\it cure\_rate}. Both are
        drawn as functions of months-past-due, {\it mpd}.}}} 
  \label{fig:1}
\end{figure}

Figure~\ref{fig:1} shows  the cure rate computed  from the survival
function compared with the raw Markov chain results.

\end{ex}

\begin{ex}\label{ex:statow} {\bf A State-Owned-Corporations Portfolio}

A portfolio of select corporate loans to state-owned corporations was 
tested for cure rate. The portfolio consists of 97 loans with total value
\euro892.8 Thousand. All measurements are taken at the end of March 2017.

The transition matrix $A$ is as follows:
\footnotesize
$$
A=\left(
\begin{array}{rrrrrrrrrr}
1 &0 &0 &0 &0 &0 &0 &0 &0 &0\\
0 &1 &0 &0 &0 &0 &0 &0 &0 &0\\
0.37 &0.63 &0 &0 &0 &0 &0 &0 &0 &0\\
0 &0 &0 &0.25 &0 &0.6 &0.15 &0 &0 &0\\
0 &0.45 &0 &0 &0.12 &0 &0.19 &0.15 &0.01 &0.08\\
0 &0 &0 &0.3 &0 &0.25 &0.45 &0 &0 &0\\
0 &0 &0 &0.4 &0 &0.37 &0.23 &0 &0 &0\\
0 &0.4 &0 &0 &0.01 &0 &0.08 &0.31 &0 &0.2\\
0 &0.21 &0 &0 &0.01 &0 &0.05 &0.11 &0.41 &0.21\\
0 &0.47 &0.01 &0 &0 &0 &0 &0.03 &0.27 &0.22\\
\end{array}
\right)
$$
\normalsize

Notice, that states $S_3$, $S_5$ and $S_6$ form a separate recurrent 
communication class and, hence, $S_5$ is not a transitive state. The
current cure-rate model is inapplicable for this
portfolio\footnote{Most banks in Bulgaria are avoiding the use cure
  rate for corporate portfolios.}. In fact we conclude  that there is
a pattern of loans cycling between these three states without ever
being 
lost or cured.
\end{ex}

\section{Portfolios optimization}\label{sec:opt}
\subsection{Cyclicity} As seen in Example~\ref{ex:statow}, it is 
possible to find a recurrent class which contains both performing and 
non-performing states. This is an indication that loans with this
risk profile may oscillate between performance and non-performance
without ever becoming cured or lost. Instead of trying
to remedy the situation using cure rate the management should seek to
optimize the contract for this type of loans.
\subsection{Hazard rate}  Example~\ref{ex:cc} shows how the  hazard
rate 
might turn out constant. 
In fact, a large part of
the tested portfolios of this type exhibited $k$ 
statistically indistinguishable from 1. 

This may be due simply to the shallowness of
the portfolio, however, it may indicate that young loans die
unnecessarily quickly. 
Therefore, it may be worth for the bank management to consider contractual
change in order to better channel the behavioral patterns of their clients.

\subsection{Time to failure or recovery}
This model allows to compute expected time $L_i$ it takes for a loan
in state $S_i$ to be resolved, i.e., to either fail, or cure. 
\begin{prp}\label{prp:2}
The time $L_i$ to absorption of a transitive state $i$, $i\ge 2$,
    can be obtained by summing the entrees of the corresponding row (i.e., the
    $(i-1)$\ts{st} row) of the matrix $(I-S)^{-1}$.
\end{prp}
\proof 
One can see that
that $L_i$ is the sum 
$$ L_i = \sum_{j=2}^N L(i,j),
$$
where $L(i,j)$ is the expected time a state spends in state $S_j$
starting initial  state $S_i$. Moreover, taking into account that
$$ (I-S)^{-1} = I + S + S^2 + \cdots
$$
 a simple computation shows, 
$L(i,j)$ is the element of that matrix which stays in $(i-1)$\ts{st}
row and 
$(j-1)$\ts{st} column.
\qed

Proposition~\ref{prp:2} produces a tool for estimation of the
expected period of uncertainty for loans. 
Furthermore, in a similar fashion one can develop a early-warning
system for prognosticating potential non-performing loans, by
computing the times $L(3,5)$ and $L(4,5)$, etc.

\section{Conclusions}
I suggest a Markov-chain model which, together with a
survival model, can be used to estimate the cure rate in a portfolio of
loans with homogeneous risk. \footnote{The model was subsequently tested by
computing cure rates for various select subportfolios of the retail product
line over a period  in years 2015-2016. The data was obtained from
three small Bulgarian banks. The analysis 
produced cure rates ranging between 3-22\%. These results appear in line
with the guidance of \cite{aqr}.}

Furthermore, we show that this technique produces the following
instruments can be of use for making these portfolios more
efficiently.
\begin{enumerate}
\item One can test if the portfolio demonstrates cyclical behavior,
  which defeats the purpose of  computation of CR. 
\item One can compute the hazard-rate function of the portfolio and
  study it for further portfolio optimization, particularly in cases
  of unexpected hazard rate. 
\item Expected time-to-resolution together with probabilities of default may
  be used for monitoring loans which are past-due over an extended
  period.
\item Expected time-to-NPL can be computed to aid the development of
  an early warning system.
\end{enumerate}

%\bigskip

\address{Institute of Mathematics and Informatics,
Bulgarian Academy of Sciences,
Acad. G. Bonchev Str., Block 8, 1113 Sofia, Bulgaria}

\email{vboutcha@math.bas.bg}

%\begin{flushright}
%\newlength{\miniwidth}
%\setlength{\miniwidth}{0.5\textwidth}
%\settowidth{\miniwidth}{\footnotesize{\textsl{Institute of Mathematics and Informatics}}}
% \begin{minipage}[t]{\miniwidth}
%   \begin{center}
%     \begin{footnotesize}
%\noindent{%
%         Institute of Mathematics and Informatics\\
%         Bulgarian Academy of Sciences\\
 %        Acad. G. Bonchev Street, Block 8\\
 %        1113 Sofia, Bulgaria\\
%         e-mail:} \tt{vboutcha@math.bas.bg}
  %   \end{footnotesize}
%  \end{center}
% \end{minipage}
%\end{flushright}

\end{document}